\documentclass{SPAICE}
\usepackage{booktabs}
\usepackage{siunitx}
\def\authorEmail{sn4533@nyu.edu}
\def\AuthorShort{S. Nikiforov}

\author[1]{Sergey Nikiforov\thanks{Corresponding author. Email: \authorEmail}}

\affil[1]{Center for Astrophysics and Space Science (CASS),
New York University Abu Dhabi,
Abu Dhabi, United Arab Emirates}

\title{Physics-Informed Multi-Task Surrogate Model for the Martian Nightside Thermosphere}

\begin{document}

\maketitle
\begin{abstract}
\indent Modeling the Martian nightside thermosphere remains challenging due to sparse in situ sampling and strong coupling among transport, magnetic, and seasonal processes. Purely data-driven models can produce non-physical artifacts, such as density inversions, in poorly sampled altitude regimes. 

We present a multi-task physics-informed neural network that simultaneously predicts the base-10 logarithmic densities of four neutral species (O, CO$_2$, N$_2$, and Ar) using more than a decade of MAVEN/NGIMS observations (MY 32--38, 2014--2025). A shared backbone learns a common representation of the nightside thermospheric state and branches into species-specific output heads.

A weak monotonicity prior is incorporated via automatic differentiation by penalizing positive vertical gradients in logarithmic density. Experiments using an orbit-disjoint train/validation/test split show that physics-informed regularization substantially reduces non-physical inversions while preserving predictive skill and slightly improving it in the best-performing configuration, as measured by RMSE, MAE, and $R^2$.

The resulting model provides a computationally efficient surrogate for nightside thermospheric reconstruction with improved vertical consistency.
\end{abstract}

\section{Introduction}

Modeling the Martian nightside thermosphere is complicated by the lack of direct local solar forcing and the dominance of complex transport processes, compounded by incomplete observational coverage. 
Global Circulation Models (GCMs), such as the LMD-GCM~\cite{Forget1999} and the NASA Ames GCM~\cite{Haberle2019}, provide a robust framework for large-scale dynamics. However, they are computationally intensive and not always practical for rapid evaluation. The resolution of complex hydrodynamics and photochemistry across a global grid requires substantial high-performance computing power. Engineering references like Mars-GRAM~\cite{Justus2001MarsGRAM} offer standardized, computationally efficient average profiles. However, capturing localized, small-scale variability due to short-lived space weather effects remains a significant challenge.

More than a decade of in situ measurements from the MAVEN mission~\cite{Jakosky2015} provides valuable insight into Martian upper-atmosphere variability. However, nightside coverage remains inherently irregular, as the mission was not optimized for systematic sampling of these regions. Such sparsity can cause purely data-driven models to extrapolate poorly, producing non-physical artifacts such as density inversions. To address this limitation, we introduce a multi-task physics-informed neural network (MT-PINN)~\cite{Raissi2019} that augments MAVEN observations with a weak monotonicity constraint on the vertical density gradient.

The model is intended as an observationally constrained complement to existing thermospheric frameworks. It is designed to improve representation in regions where global simulations~\cite{Lewis1999} and empirical climatologies~\cite{Millour2018} can deviate from in situ measurements~\cite{Bougher2015}, particularly under sparsely sampled nightside conditions. 

MT-PINN is not intended to replace physics-based GCMs. It provides a fast observational surrogate that preserves a simple constraint on vertical structure and can be queried in regimes where data coverage is limited.

A fast nightside density surrogate can also support mission-oriented workflows, including profile reconstruction along arbitrary trajectories, sensitivity studies across seasons and space weather conditions, and model--data comparison without repeated GCM runs.

MT-PINN therefore provides a direct link between irregular in situ observations and physics-based model comparison. Its differentiable output also allows profile-by-profile comparison with GCM results under matched environmental conditions.

\section{Data and Parameter Formulation}

We analyze MAVEN Key Parameters (KP)~\cite{MAVEN_KP} from late MY 32 to early MY 38, spanning a complete solar cycle and the MY 34 global dust storm~\cite{Sanchez2019,Stone2022}. The broader dataset combines in situ measurements from NGIMS~\cite{mahaffy2015} (neutral densities), MAG~\cite{Connerney2015} (magnetic fields), SWEA~\cite{Mitchell2016} and SEP~\cite{Larson2015} (particle fluxes), along with geometry from SPICE~\cite{ACTON2018}. Atmospheric dust loading is parameterized using reconstructed climatology maps~\cite{Montabone2015,Montabone2020}.

The input space $\mathbf{X}$ encompasses spatial, temporal, and geophysical drivers. In addition to standard kinematic data (altitude, latitude, longitude, solar zenith angle, local solar time) and seasonal parameters $(L_s)$, we explicitly include external energy inputs that influence nightside thermospheric dynamics. When direct solar extreme ultraviolet insolation is absent at the local nightside, the thermospheric state is shaped by global day-to-night circulation and localized heating from particle precipitation. 
To represent upstream solar wind conditions when direct local measurements are unavailable, we use density, velocity, temperature, and dynamic pressure estimates from an external Gaussian-process solar wind model~\cite{Azari2024}. Nightside conditions are identified by solar zenith angle exceeding $90^\circ$.

The model is trained jointly for four neutral species (O, CO$_2$, N$_2$, and Ar) using a multi-task learning framework. Measurements with non-positive density or reported density error above the species-specific 99th percentile are excluded from the corresponding target mask. To account for varying data fidelity, the remaining NGIMS measurements are weighted according to their reported density error values, as described in Section~3.2.

\section{Multi-Task Physics-Informed Neural Network}
\subsection{Neural Network Architecture}

We adopt a multi-task learning framework~\cite{Caruana1997, Crawshaw2020} combined with a physics-informed neural network formulation~\cite{Raissi2019, Karniadakis2021}. The model simultaneously predicts the base-10 logarithms of neutral number densities for four thermospheric species: O, CO$_2$, N$_2$, and Ar.

For each species $i$, we define the target variable as
\begin{equation}
y_i = \log_{10}(\rho_i)
\end{equation}
where $\rho_i$ denotes the measured neutral number density.
The network produces corresponding predictions
\begin{equation}
\hat{y}_i = \log_{10}(\hat{\rho}_i)
\end{equation}
with $\hat{\rho}_i$ representing the model-estimated density.
All regression is therefore performed in log-density space.

Joint prediction is motivated by the fact that these species share a common background neutral temperature and large-scale wind fields, which govern their transport and diffusive separation under external drivers such as solar wind forcing and crustal magnetic topology.

The architecture consists of a shared feature extractor (backbone)
followed by species-specific linear output heads.
The shared backbone maps the inputs to a latent representation that captures common responses to altitude, solar zenith angle, $L_s$, magnetic structure, and particle precipitation.

The backbone is implemented as a three-layer multilayer perceptron
(Table~\ref{tab:arch}):

\begin{table}[!h]
\centering
\caption{Multi-Task PINN architecture.}
\label{tab:arch}
\begin{tabular}{ll}
\hline
Backbone & $d_{in}\!\rightarrow\!256\!\rightarrow\!128\!\rightarrow\!64$ \\
         & BN + ReLU (+DO for first two layers) \\
Heads    & 4 $\times$ Linear(64,1) \\
\hline
\end{tabular}

\vspace{2pt}
{\small BN: batch normalization; ReLU: rectified linear unit; DO: dropout}
\end{table}

Each species-specific head produces
\begin{equation}
\hat{y}_i = W_i h + b_i,
\quad i \in \{ \mathrm{O}, \mathrm{CO_2}, \mathrm{N_2}, \mathrm{Ar} \},
\end{equation}
where $h \in \mathbb{R}^{64}$ denotes the shared latent feature vector.
The final prediction vector is formed by concatenation:
\begin{equation}
\hat{\mathbf{y}} =
\mathrm{concat}(\hat{y}_{\mathrm{O}},
\hat{y}_{\mathrm{CO_2}},
\hat{y}_{\mathrm{N_2}},
\hat{y}_{\mathrm{Ar}})
\end{equation}

Modeling densities in logarithmic space improves numerical stability and effectively captures the approximately exponential decrease of density with altitude. It also reduces variability between species, whose absolute densities differ by several orders of magnitude in the 140--350~km regime.

\subsection{Loss Function}
To train the MT-PINN, we formulate a composite objective that balances fidelity to in situ observations with a physically motivated structural prior. The total loss is defined as:
\begin{equation}
\mathcal{L}_{\text{total}} =
\mathcal{L}_{\text{data}} +
\lambda \mathcal{L}_{\text{phys}},
\end{equation}
where $\mathcal{L}_{\text{data}}$ enforces agreement with measurements,
and $\mathcal{L}_{\text{phys}}$ introduces a weak monotonicity constraint
in the vertical direction. Here, physics-informed regularization refers to a structural constraint rather than direct enforcement of a governing partial differential equation.

Thermospheric measurements exhibit varying signal-to-noise ratios across altitude,
species, and environmental conditions. To account for this heterogeneity, we employ an uncertainty-weighted mean squared error:
\begin{equation}
\mathcal{L}_{\text{data}} = \frac{1}{N} \sum_{j=1}^{N} \frac{\sum_{i} m_{j,i} w_{j,i} (\hat{y}_{j,i} - y_{j,i})^2}{\sum_{i} m_{j,i} + \epsilon}
\end{equation}
where $N$ is the batch size, $j$ indexes the individual samples within a batch, and $i$ indexes the gas species. The variable $m_{j,i}$ denotes a validity mask (1 for a valid NGIMS measurement and 0 otherwise), and $y_{j,i}$ ($\hat{y}_{j,i}$) represents the true (predicted) log density. The measurement weights are defined as
\begin{equation}
\tilde{w}_{j,i} = \frac{1}{u_{j,i}+0.05},
\qquad
w_{j,i} =
\frac{\tilde{w}_{j,i}}
{\langle \tilde{w}_{i} \rangle},
\end{equation}
where $u_{j,i}$ denotes the reported NGIMS density error quantity and $\langle \tilde{w}_{i} \rangle$ is the mean weight for valid measurements of species $i$. The validity mask follows the species-specific filtering described in Section~2. A small constant $\epsilon$ is added to the denominator for numerical stability. The per-sample normalization prevents samples containing multiple valid species from contributing disproportionately to the batch loss.

Under approximately hydrostatic conditions, neutral densities decrease roughly exponentially with altitude. Purely data-driven regressors may produce non-physical positive vertical gradients (density inversions), particularly in sparsely sampled regimes.

To mitigate this behavior, we penalize positive vertical gradients in the logarithmic density:
\begin{equation}
\mathcal{L}_{\text{phys}} =
\frac{1}{T}
\sum_{i=1}^{T}
\left(
\frac{1}{N} \sum_{j=1}^{N}
\left[
\mathrm{ReLU}
\left(
\frac{\partial \hat{y}_{j,i}}{\partial z_{j}}
\right)
\right]^2
\right)
\end{equation}
where $T=4$ is the number of species, $N$ is the batch size, and $z_{j}$ denotes the altitude corresponding to sample $j$. The $\mathrm{ReLU}$ operator selectively penalizes only positive gradients, leaving negative gradients unconstrained.

Gradients are computed via automatic differentiation with respect to the normalized altitude feature. To recover dimensional consistency, we rescale the gradients by the altitude standard deviation used during input normalization.

This soft regularization does not enforce strict hydrostatic balance. Instead, it reduces large-scale non-physical inversions while allowing localized deviations due to transient heating or measurement noise. The hyperparameter $\lambda$ controls the trade-off between observational accuracy and physical consistency.

\subsection{Training and Data Splitting Strategy}

To ensure a realistic evaluation of model generalization, we perform data partitioning at the orbit level rather than at the individual sample level. All measurements belonging to a single MAVEN orbit
are treated as a coherent block and assigned entirely to either the training, validation, or test subset.

Successive measurements along a single orbit are strongly correlated in altitude, latitude, magnetic topology, and environmental conditions.
Random sample-level shuffling would therefore introduce data leakage, allowing the network to interpolate between nearly identical adjacent points rather than learning the underlying relationships
between geophysical drivers and thermospheric structure. Such leakage typically results in artificially inflated performance metrics and overestimates of predictive skill.

We adopt a strict orbit-disjoint three-way split. Approximately 80\% of the data are assigned to training, while the remaining 20\% are divided evenly into validation and test subsets. The validation subset is used for early stopping, learning-rate scheduling, and model checkpoint selection. The held-out test subset is used for final performance reporting.

The resulting split tests generalization to independent nightside trajectories with orbital conditions not used during training, including variations in crustal magnetic field configuration, season ($L_s$), and upstream solar wind forcing.

\section{Results}

We evaluate MT-PINN on two criteria: generalization to unseen orbital passes and consistency of the predicted vertical structure. Unless stated otherwise, all reported performance metrics refer to the held-out test set.

\subsection{Calibration of Physical Regularization}

Rather than treating the physics weight $\lambda$ as a fixed hyperparameter, we interpret it as a calibration parameter controlling the balance between statistical fidelity and physical structure. We therefore explore a range:
\begin{equation}
\lambda \in \{0,\ 0.01,\ 0.05,\ 0.10,\ 0.20,\ 0.50,\ 0.75,\ 1.00\}.
\end{equation}

Predictive skill is evaluated in base-10 log-density space using RMSE, MAE, and $R^2$, computed only for valid MAVEN/NGIMS measurements of each species. Physical consistency is quantified via the inversion rate, defined as the fraction of held-out test inputs exhibiting a locally positive vertical gradient:
\begin{equation}
\frac{\partial \hat{y}_i}{\partial z} > 0.
\end{equation}
Because the surrogate predicts all four species at every input state, inversion rates are evaluated over all held-out test inputs.

\begin{figure}[t]
    \centering
    \includegraphics[width=.95\columnwidth]{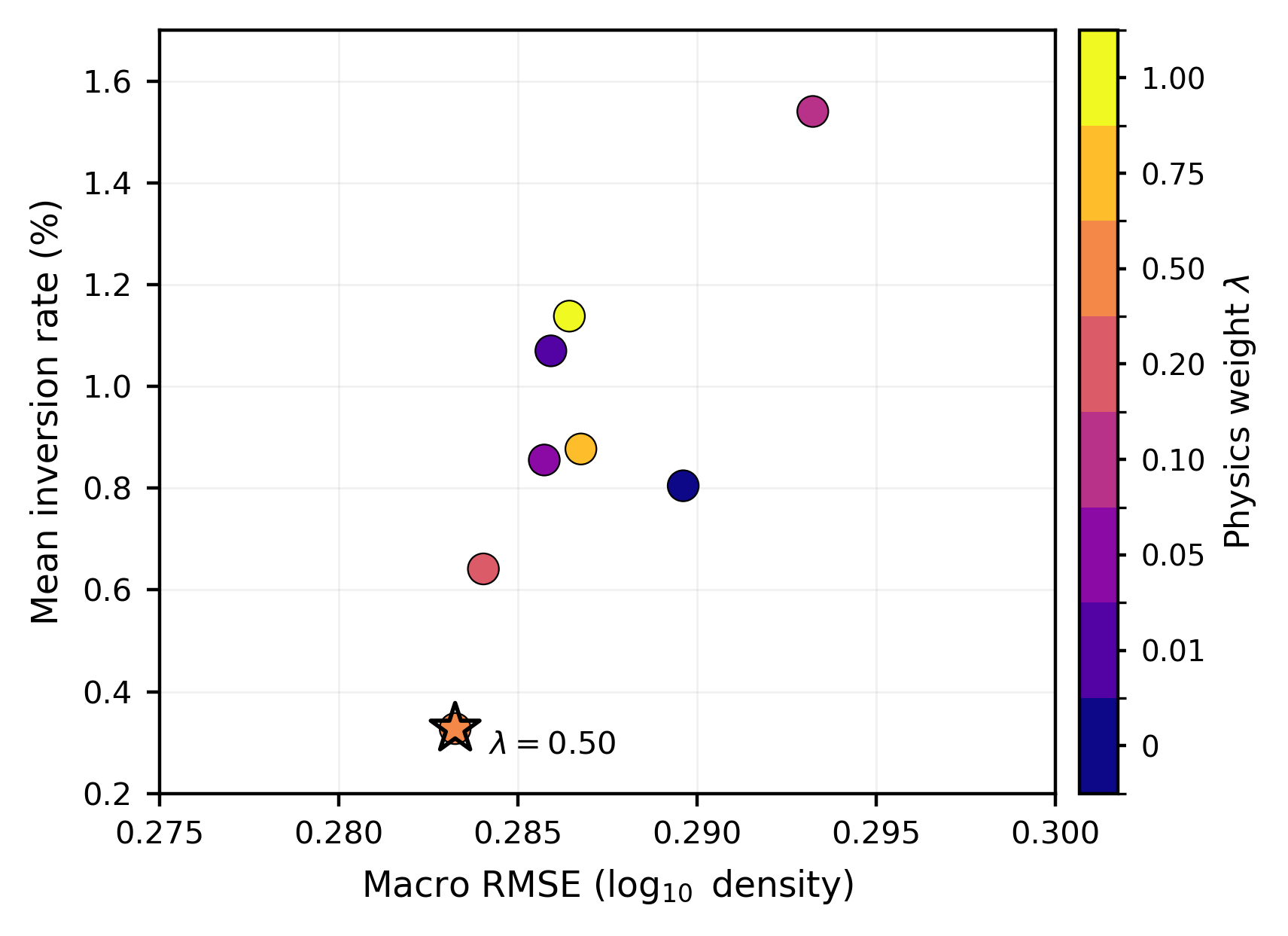}
    \caption{Accuracy--consistency trade-off on the held-out test set.
    Macro-averaged RMSE across O, CO$_2$, N$_2$, and Ar is shown against
    the mean inversion rate across the four species. Marker color denotes
    the tested physics regularization weight $\lambda$. Selected values
    are annotated for reference, and the star highlights $\lambda=0.50$.}
    \label{fig:tradeoff}
\end{figure}

Figure~\ref{fig:tradeoff} illustrates the accuracy--consistency
trade-off across the tested regularization weights. Among the tested values, $\lambda=0.50$ gives both the lowest macro-averaged RMSE and the lowest mean inversion rate across species.

Table~\ref{tab:test_tradeoff} summarizes the resulting behavior. Purely data-driven training ($\lambda=0$) already provides strong predictive skill but exhibits residual non-physical vertical gradients. The response to regularization is not strictly monotonic across the tested range or across individual species. At $\lambda=0.50$, the model achieves the lowest macro-averaged RMSE and MAE and the highest $R^2$, while substantially reducing inversion rates for O and N$_2$ and maintaining a very low inversion rate for Ar. Increasing $\lambda$ to 0.75 or 1.00 worsens predictive performance and increases inversion rates relative to $\lambda=0.50$.

\begin{table}[!h]
\centering
\caption{Held-out test-set performance as a function of the physics regularization weight. Predictive metrics are macro-averaged across O, CO$_2$, N$_2$, and Ar. Inversion rates are evaluated over all held-out test inputs and are reported in percent. Best values in each column are shown in bold.}
\label{tab:test_tradeoff}
\scriptsize
\setlength{\tabcolsep}{3.5pt}

\begin{tabular}{@{}cccc@{}}
\toprule
$\lambda$ & RMSE & MAE & $R^2$ \\
\midrule
0.00 & 0.2896 & 0.2179 & 0.9231 \\
0.01 & 0.2859 & 0.2152 & 0.9255 \\
0.05 & 0.2857 & 0.2140 & 0.9254 \\
0.10 & 0.2932 & 0.2224 & 0.9213 \\
0.20 & 0.2840 & 0.2134 & 0.9262 \\
0.50 & \textbf{0.2833} & \textbf{0.2124} & \textbf{0.9268} \\
0.75 & 0.2868 & 0.2154 & 0.9249 \\
1.00 & 0.2864 & 0.2157 & 0.9248 \\
\bottomrule
\end{tabular}

\vspace{3pt}

\begin{tabular}{@{}ccccc@{}}
\toprule
$\lambda$ & O Inv. & CO$_2$ Inv. & N$_2$ Inv. & Ar Inv. \\
 & (\%) & (\%) & (\%) & (\%) \\
\midrule
0.00 & 2.226 & \textbf{0.533} & 0.456 & 0.0036 \\
0.01 & 2.918 & 0.716 & 0.630 & 0.0143 \\
0.05 & 2.039 & 0.753 & 0.613 & 0.0152 \\
0.10 & 4.122 & 1.122 & 0.876 & 0.0411 \\
0.20 & 1.433 & 0.950 & 0.182 & \textbf{0.0006} \\
0.50 & \textbf{0.608} & 0.586 & \textbf{0.115} & 0.0009 \\
0.75 & 2.499 & 0.610 & 0.393 & 0.0062 \\
1.00 & 2.806 & 0.858 & 0.867 & 0.0214 \\
\bottomrule
\end{tabular}
\end{table}

An important practical advantage of MT-PINN is that it produces continuous, differentiable vertical profiles. This enables direct diagnostics of physically interpretable quantities such as $\partial \hat{y}_i/\partial z$ and supports integration with physics-based workflows. In this sense, $\lambda$ controls not only statistical fit, but also the reliability of gradient-based interpretation in sparsely sampled regimes.

\subsection{Contextual Baseline Comparison}

As a non-neural reference point, we trained an XGBoost regressor to predict log density using a compact feature set with engineered cyclical variables and an orbit-disjoint split (every fifth orbit assigned to test). For O, the tuned model achieves RMSE $\approx 0.207$ and $R^2 \approx 0.911$ in logarithmic density space. Because this baseline uses a reduced feature set and a different split protocol than the MT-PINN experiments, we treat it as contextual benchmarking rather than a strictly controlled head-to-head comparison.

While tree-based ensembles can yield strong pointwise predictive metrics, they do not naturally provide smooth, differentiable vertical structure. As a result, gradient-based diagnostics ($\partial \hat{y}/\partial z$) are either undefined or dominated by piecewise constant behavior. For atmospheric applications where vertical gradients carry physical meaning and are used in downstream coupling and consistency checks, this motivates the use of differentiable models such as MT-PINN.

\section{Discussion and Outlook}

\subsection{Limitations and Scope}

The model is restricted to nightside neutral densities within the 140--350~km altitude range and is not a globally self-consistent thermospheric model. The weak monotonicity constraint improves vertical plausibility but does not enforce hydrostatic balance, energy conservation, or self-consistent momentum coupling.

In a multi-task setting, shared latent representations may propagate systematic biases between species. Their specific contribution to multi-species prediction requires further controlled investigation.

Localized departures from monotonicity may also arise from gravity waves, transient heating, or measurement noise. The regularization is therefore intentionally soft, reducing large-scale non-physical inversions without enforcing strictly monotonic profiles.

\subsection{Computational Efficiency and Model Intercomparison}

A systematic quantitative intercomparison with classical thermospheric models remains outside the scope of the present study. Once trained, the network provides rapid pointwise evaluation without the computational cost of running a full three-dimensional circulation model.

The weak structural constraint adds physical guidance without changing the model into a full dynamical solver. This is useful when rapid evaluation is more important than a globally self-consistent atmospheric solution.

\subsection{Implications}

For the Martian nightside, sparse and irregular sampling makes purely empirical reconstruction difficult. The present results show that even a weak structural prior can reduce non-physical vertical gradients without reducing predictive skill.

In this model, a shared multi-species representation can be combined with weak physical regularization while retaining accurate density predictions and improving vertical consistency.

\subsection{Future Roadmap}

A systematic intercomparison with GCM outputs across seasons, crustal magnetic configurations, and extreme solar events is ongoing. Future work will test temperature-dependent hydrostatic constraints based on species-specific scale heights and extend the model to ionospheric constituents.

\begin{acknowledgments}
   This material is based upon work supported by Tamkeen under the NYU Abu Dhabi Research Institute grant CASS.
\end{acknowledgments}

\printbibliography
\addcontentsline{toc}{section}{References}

\end{document}